\documentclass{article}

\usepackage{arxiv}

\usepackage[utf8]{inputenc} 
\usepackage[T1]{fontenc}    
\usepackage{hyperref}       
\usepackage{url}            
\usepackage{booktabs}       
\usepackage{amsfonts}       
\usepackage{nicefrac}       
\usepackage{microtype}      
\usepackage{cleveref}       
\usepackage{lipsum}         
\usepackage{graphicx}
\usepackage{doi}
\usepackage{amsthm}
\usepackage[english,russian]{babel}
\usepackage[square,sort,comma,numbers]{natbib}

\newtheorem{definition}{Definition}[section]
\newtheorem{theorem}{Theorem}[section]

\title{Real-time Regular Expression Matching}

\author{ \href{https://orcid.org/0000-0002-8835-6583}{\includegraphics[scale=0.06]{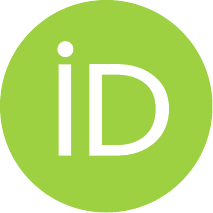}\hspace{1mm}Alexandra~Bernadotte$^{1,2,3,4}$}\thanks{Use footnote for providing further
		information about author (webpage, alternative
		address)---\emph{not} for acknowledging funding agencies.} \\
$^{1}$ \quad  Faculty of Mechanics and Mathematics,	Lomonosov Moscow State University\\
	Leninskiye~Gory~1, Moscow,  119991, Russia;\\
$^{2}$ \quad  National University of Science and Technology MISIS \\
	Leninskiy~Prospekt~4, Moscow,  119049, Russia;\\
$^{3}$ \quad  Neurosputnik LLC,	Vernadskogo~Prospekt~96, Moscow,  119049, Russia;\\
$^{4}$ \quad  Rebis LLC, Vernadskogo~Prospekt~96, Moscow,  119049, Russia;\\
	\texttt{bernadotte.alexandra@intsys.msu.ru} \\
		\texttt{bernadotte@neurosputnik.com} \\
}

\renewcommand{\shorttitle}{\textit{arXiv} Template}

\hypersetup{
pdftitle={Real-time Regular Expression Matching},
pdfsubject={math, cs},
pdfauthor={Alexandra~Bernadotte},
pdfkeywords={Regular Expression, Regular Expression Matching, Regular expression pattern matching, regular expression denial-of-service, Real-time Matching, Automata theory, ReDoS, finite-state machine},
}

\begin{document}
\maketitle

\begin{abstract}
This paper is devoted to finite state automata, regular expression matching, pattern recognition, and the exponential blow-up problem, which is the growing complexity of automata exponentially depending on regular expression length. 
	This paper presents a theoretical and hardware solution to the exponential blow-up problem for some complicated classes of regular languages, which caused severe limitations in Network Intrusion Detection Systems work. 
	The article supports the solution with theorems on correctness and complexity. 
\end{abstract}

\keywords{Regular Expression \and Regular Expression Matching \and Real-time Matching \and Automata theory \and telesurgery \and NIDS \and cybersecurity }

\section{Introduction}

Regular expression matching is a search for patterns in discrete data using regular expressions. Regular expression matching is used in cybersecurity tasks, computational biology for interacting sequencing searching, and at any software and hardware analysers of a signal represented by a sequence of symbols or patterns~\cite{Kumar.1, Kumar.2, Kumar.3, Antonova.1, Antonova.2, Alex.2, Podkolzin.1}.

There are tasks for which pattern detection using regular expression matching should be accurate, fast, and carried out on-line. For example, such tasks include the cybersecurity of telesurgical operations, in which surgeons remotely control the operating robotic device~\cite{Tariq.1, Alex.5, Alex.6}. 

The following cyber attacks are typical for the above application: intentional manipulation, intentional modification of control commands, and denial of service. 

The good news is that signs of the above attacks can be detected in network traffic. Cybersecurity experts compile regular expression databases of characteristic features of malicious code, XSS, Trojan signs, etc. Existing Network Intrusion Detection Systems (NIDS) scan network traffic for malware using these regular expression databases such as Bro and Snort.

These NIDS often use finite state automata (FSA) to implement fast traffic analysis. State machines are an efficient and fast way to parse regular expressions.

However, there is a mathematical problem of an exponentially growing number of states of a finite automaton from the length of regular expressions specific to a certain type of regular expression. The problem of the exponentially growing number of states of the state machine on the length of regular expressions leads to the impossibility of implementing compact hardware for analysing malicious traffic in practice since the memory requirements for storing a state machine that uses existing regular expressions from the Bro and Snort databases, exceed reasonable limits. From a practical point of view, analysing traffic for malware on-line is impossible using classical FSA since the solution has to be either fast or space efficient.

FSA is the best thought for pattern recognition in NIDS and data mining. Space is usually a more critical issue than time. However, for NIDS, both parameters are essential.

The use of this problem in intrusion detection systems is called The Regular expression Denial of Service (ReDoS). The Regular expression Denial of Service (ReDoS) is a Denial of Service attack that exploits a finite automaton's exponentially growing number of states and makes NIDS works slowly (exponentially related to input size). An attacker can then cause a program using a Regular Expression to enter these extreme situations and then hang remote hardware for a very long time~\cite{owasp}.

This work is devoted to this problem and its solution. This paper provides the solution for the exponential problem of most regular expressions, which caused severe limitations in NIDS work.

\subsection{Definitions}
\subsubsection{Regular languages and regular expressions}
\begin{definition}A regular language (event or set) over the alphabet $A$ is defined recursively as follows~\cite{Kudryavcev.1, Kudryavcev.2}.:
\begin {enumerate}
\item $\emptyset, \{\lambda\}$ and $\{a\}$ are regular languages, where $\lambda$ is an empty word, $a \in A$;
\item if $M_1$ and $M_2$~ are regular languages, then $M_1 \cup M_2$ union,   $M_1 \cdot M_2$ concatenation, and Kleene star $M_1^*$ are regular languages, where

$$
\begin {array}{c}
M_1 \cup M_2 = \{ \alpha_1\cup\alpha_2 | \alpha_1 \in M_1, \alpha_2 \in M_2\},\\
M_1 \cdot M_2 = \{ \alpha_1\cdot\alpha_2 | \alpha_1 \in M_1, \alpha_2 \in M_2\},\\
M_1^* = \{\lambda\}\cup\{\alpha_1\cdot \dots \alpha_n |\alpha_i \in M_1, i \in \overline{1,n}, n \in N\}.
\end {array}
$$
\end {enumerate}
\end{definition}

Regular languages correspond to regular expressions. 

\begin{definition} Regular expression is defined recursively as follows:
\begin {enumerate}
\item $\emptyset, \lambda, a$ are regular expressions, where $a \in A$;
\item if $R_1$, $R_2$ are regular expressions, then the $(R_1\cup R_2), (R_1\cdot R_2)$, and $(R_1)^{*}$ are regular expressions.
\end{enumerate}
\end{definition}

Regular expression is a language over the alphabet $A\cup\{\cup, (, ), \cdot, *\}$, where the operation ``$*$'' has the maximum priority and the operation ``$\cup$'' has the minimum. 

The syntax and semantics of the regular expressions that are supported by PCRE (Perl Compatible Regular Expressions) are described below: 

In practice, in the Snort, Zeek, and Cisco databases, regular expressions
are represented using the PCRE--notation  (Perl Compatible Regular Expressions)~\cite{Perl.1}. Here is the PCRE--notation that we will use in the paper:
\begin {itemize} 
\item [] ``.''  is any character;
\item [] ``.*'' is 0 or more quantifier;
\item [] ``+'' is 1 or more quantifier;
\item [] ``$\{n\}$'' is $n$ quantifier ;
\item [] ``$\{n, m\}$'' is form $n$ to $m$ quantifier;
\item [] ``$|$'' is logic ``or'';
\item [] ``$\neg$'' is logic ``not'';
\item [] ``$[$'' is POSIX character class (only if followed by POSIX syntax);
\item [] ``$]$'' is terminates the character class;
\item [] ``$[0-9]$''is any decimal digit; 
\item [] ``$[A-Za-z0-9]$''is any decimal digit or letter character.
\end {itemize}

\subsubsection{Finite state automata}

Finite state automata (FSA), also known as finite state machines (FSM), are used to recognize regular languages by accepting or rejecting a given string (word) of a certain regular language. 

\begin{definition}
A FSA is a tuple, $(A, Q, F, \varphi)$, where $A$ is a finite set of input symbols called the input alphabet, $Q$ is a finite set of states (state alphabet), $F$ is a set of accept states $F \subseteq Q$, $\varphi$ is a transition function, $\varphi : Q\times A \to Q$.
\end{definition}

FSAs can be subdivided into some classes. A FSA with a starting state called the initial finite automaton, and it is a 5-tuple $(A, Q, B, \varphi, \psi, q_0)$, where $q_0$ is an initial state in which the FSA starts~\cite{Kudryavcev.1, Kudryavcev.2}.

A finite state transducer (FST), following the terminology for Turing machines, has input and output tape. Instead of accepting states, this class of FSAs has an output alphabet.

\begin{definition}
A FST is a tuple, $(A, Q, B, \varphi, \psi)$, where $A$ is a finite set of input symbols called the input alphabet, $Q$ is a finite set of states (state alphabet), $B$ is a set of accept states $B \subseteq Q$, $\varphi$ is a transition function, $\varphi : Q\times A \to Q$, $\psi$ is an output function, $\psi : Q\times A \to B$.
\end{definition}

FSAs recognize regular languages. Acceptors (detectors or recognizers) produce binary output, indicating whether or not the received input is accepted. Each state of an acceptor is either accepting or non-accepting.

According to Kleene's theorem, an event (word, string) over the alphabet $A$ is representable by a finite automaton if and only if it is a regular language~\cite{Hap.1, Rabin.1, Kudryavcev.1, Kudryavcev.2}.

Both representations of finite automata FSA and FST are equivalent. 
The FSA recognizes a regular language by mapping strings (words of the language) into the set of accepting states. In contrast, an equivalent FST recognizes a regular language by mapping strings into the output set $\{0,1\}$.


\subsubsection{Deterministic and non-deterministic finite state automata}

FSAs are classified as deterministic (DFA) or non-deterministic (NFA). 

The no-ndeterministic finite automaton (NFA) allows jumping over the empty word. It is a set $(A, Q, B, \varphi)$, where $A, Q, B$ are finite sets: input alphabet, state alphabet, and a set of accepting states $B \subseteq Q$, $\varphi$ is a transition function, $\varphi : Q\times \{A\cup \lambda \} \to 2^Q$, $\varphi$ returns a set of states.

A deterministic finite automaton (DFA) has no $\lambda$--transitions, the transitions on the empty word, and all DFA transitions are uniquely defined~\cite{Hap.1, Rabin.1, Kudryavcev.1, Kudryavcev.2}.

According to the theorem on the equivalence of deterministic and nondeterministic finite automata, any language is accepted by some NFA if and only if this language is accepted by some DFA~\cite{Hap.1, Rabin.1, Kudryavcev.1, Kudryavcev.2}: 

\begin{theorem}Let language $L \subseteq A^{*}$, and suppose $L$ is accepted by NFA $N = (A, Q, q_0, F, \delta)$. There exists a DFA $D= (A, Q^{'}, q^{'}_0, F^{'}, \delta^{'})$ that also accepts $L$. $L(N) = L(D)$.
\end{theorem}

In addition, there is an algorithm allows to convert any DFA to an equivalent NFA~\cite{Hap.1, Rabin.1}. 

Both types of automata (DFA and NFA) are used in regular expression real-time matching. In the choice of a FSA, time and space complexity are significant.

\subsubsection{Deterministic and non-deterministic complexity}
The theory of complexity of formal languages posts the question of the optimality of a regular language representation. As applied to regular languages, the classical measures of descriptive complexity are deterministic and non-deterministic complexity~\cite{Meyer, Salomaa, Brookshear, Chamberlain.1}.

\begin{definition}
The deterministic state complexity (or deterministic complexity or spatial complexity) of a regular language $L$ is the number of states of the minimal DFA recognizing $L$ (the amount of memory).
\end{definition}

\begin{definition}
The non-deterministic state complexity of a regular language $L$ is the smallest number of transitions of a NFA recognizing $L$. 
\end{definition}

\begin{definition}
The non-deterministic transition complexity (or time complexity) of a regular language $L$ is the smallest number of states of a NFA recognizing $L$ (corresponding to the clock rate). 
\end{definition}

The transition complexity gives NFAs a more realistic complexity measure than the number of states because the number of transitions determines the size of a NFA. It is generally accepted that the calculation of the value of the transition function of an automaton has constant complexity. It comes down to retrieving the value from memory.

While DFAs can be efficiently minimized, the minimization of NFAs is known to be PSPACE-complete~\cite{Jiang}. NFAs have optimal spatial complexity, but time complexity can be insufficient for application. The NFAs spatial complexity is linear in regex length. At the same time, the processing time of one character of the input word is generally also linear in the length of the regular expression.

DFAs have optimal time complexity, but the number of states (spatial complexity) can grow exponentially with the length of the regular expression. 

When passing from a NFA to a DFA, the time complexity of processing the input symbol becomes constant; on the other hand, according to Lupanov's theorem, if the input alphabet contains at least two symbols, the power of the set of states, in general, exponentially increases~\cite{Lupanov.1, Kudryavcev.1, Kudryavcev.2}.

\subsubsection{Exponential blow-up problem}

In applications with real-time regular matching, there is a problem of the exponentially growing DFA states. The exponentially growing accepting DFA states number depending on the length of the regular expression is called the exponential blow-up problem~\cite{Lupanov.1, Kudryavcev.1, Kudryavcev.2,  Maslov1, Maslov2, Salomaa}.

There are signs that cause the exponential blow-up problem: ``$.+$'', ``$.*$'', ``$.\{n\}$ '', ``$[ \neg a_i]\{n\}$, ``$.\{n, m\}$ '', and ``$[ \neg a_i]\{n, m\}$'', where $a_i$ is a word, $n, m \in \mathbb{N}$~\cite{Indyk, Maslov1, Maslov2, Salomaa}.

Special attention is paid to the exponential blow-up problem for the $\bigcup_{i=1}^{n}.*\alpha_{i}.*\beta_{i}.*$ language, where $\alpha_{i}, \beta_{i }$ are non-empty words, and  $\bigcup_{i=1}^{n}.*R_{i}.*R_{i^{'}}.*$ language, where $R_i, R_{i^{'}}$ are regular excretions. It is known that, in the general case, this regular language requires an exponential (in $n$) number of states of a DFA~\cite{Al3.9.3}.

\begin{figure}
	\centering
	\fbox{\includegraphics[width=400pt]{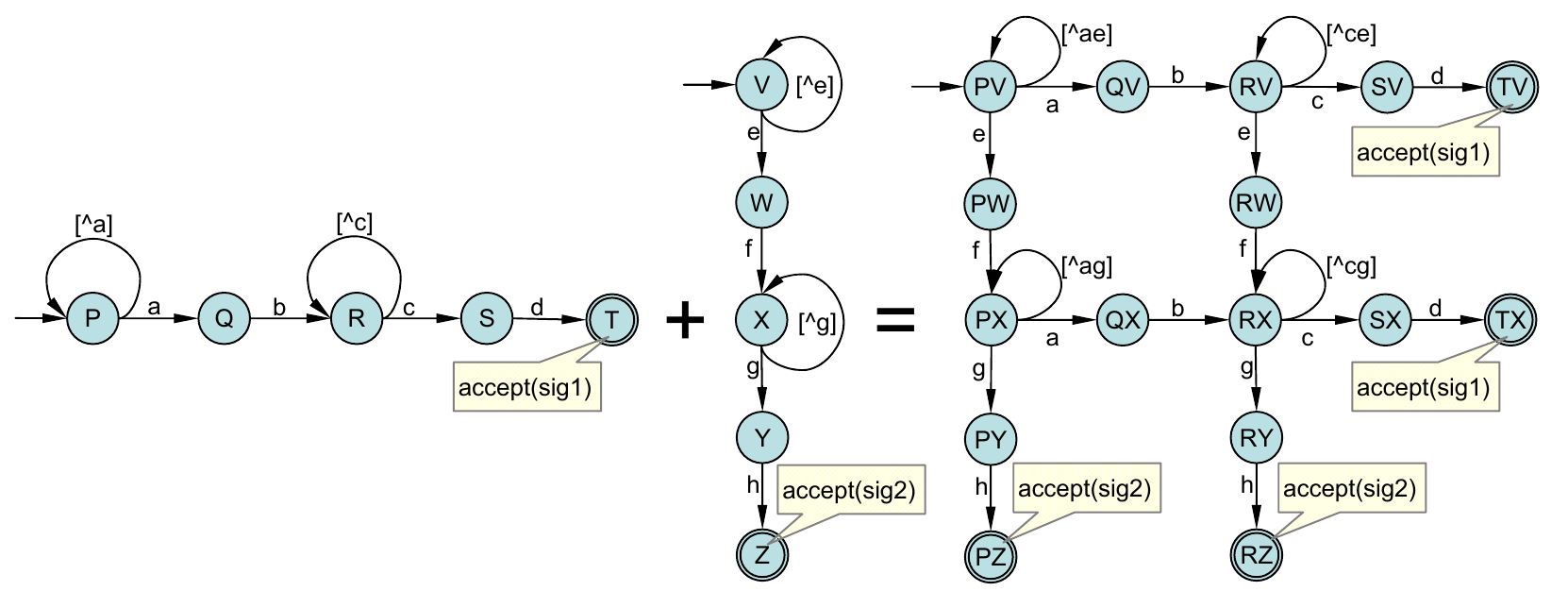}} 
	\caption{The exponential blow-up problem of the $\{(.*ab .*cd) \cup (.*ef .*gh)\}$ regular language over the alphabet $\{a, b, c, d, e, f, g, h\}$.  Source: ~\cite{Smith.1}}
	\label{fig:fig1}
\end{figure}

An illustrative example demonstrating the exponential blow-up problem for the $\bigcup_{i=1}^{n}.*\alpha_{i}.*\beta_{i}.*$ language is presented by colleagues (Fig.~\ref{fig:fig1})~\cite{Smith.1}.

It is known that regular languages can be expressed with different regular expressions. For example, the language $L$ over the finite alphabet $A, A = \{a, b,... , z, A, B, ..., Z \}$ can be written like this regular expression $R_1 = .\{10, 20\}$ or like this one $R_2 = [a|b| ... |z|A|B| ... |Z]\{10, 20\}$ or like this one $R_2 = [a-Z]\{10, 20\}$. However, repetitions such as the Kleene star or $\{k, m\}, \{k\}$ in PCRE--notation are valid as signs that cause the exponential blow-up.

\subsection{Exponential Blow-up Problem Solving Approaches Review }

There are three main approaches to solving this problem using FSA: regular language restriction specified by experts, regular language, and FSA modification.

\subsubsection{Regular language restriction}
The regular language restriction method involves the expert selection of regular expressions, which reduces the number of states of a given FSA. The disadvantage of this approach is low automation and labour intensity. 

\subsubsection{Regular language modification Approaches}
Regular language modification. This method reduces FSA complexity, which is significant from a practical point of view~\cite{Al2.9.2}.
However, this approach results in Type I (false-positive) errors, the so-called regular language extension problem, or Type II (false-negative) errors in language matching.

For example, the Exponential Blow-up Problem arises when regular expressions are combined as follows: $\bigcup_{i=1}^{n} .*R_{2i-1} .* R_{2i} .*, i \in \overline{1, n}$~\cite{Al1.9.1, Al2.9.2}. To reduce the states number of the FSA recognizing $ \bigcup .*R_i .* R_j .*, i,j \in 1, n$ language Aleksandrov~\cite{Al3.9.3} proposes to extend the regular language $\bigcup_{i=1}^{n} .*R_{2i-1} .* R_{2i} .*$ to the language  $(.*(R_1 | R_3) .* (R_2 | R_4) .* ) \cup (\bigcup_{i=3}^{n} .*R_{2i-1} .* R_{2i} .*)$. The new FSA recognizing $(.*(R_1 | R_3) .* (R_2 | R_4) .* ) \cup (\bigcup_{i=3}^{n} .*R_{2i-1} .* R_{2i} .*)$ may accept words not from the language $\bigcup_{i=1}^{n} .*R_{2i-1} .* R_{2i} .*$. However, it will have no more, and in some cases noticeably fewer states than an FSA that accepts the language $\bigcup_{i=1}^{n} .*R_{2i-1} .* R_{2i} .*$~\cite{Al1.9.1, Al2.9.2, Al3.9.3}.

\subsubsection{FSA modification}
The FSA modification approach can be implemented as FSA compression approaches 
and FSA Structural modifications~\cite{Alex.3, Alex.4}.

\subsubsection{FSA Compression Approaches}
Most compression methods are based on adding non-deterministic transitions to a DFA that recognizes a given language~\cite{Alex.4}.

In 1975, Alfred Aho and Margaret Corasick developed an algorithm that searches for words over the alphabet $A$ from a given dictionary in a string. The algorithm builds a modified initial DFA based on a prefix tree~\cite{Aho.1, Briandais}. Compression is done through the use of ``error jumps'' that are not present in the pre-built prefix tree. For each state, it is necessary to store information about only two outgoing transitions: about a deterministic transition along the corresponding symbol from the alphabet $A$ (the transition corresponds to the transition of the prefix tree) and about the transition in the case ``otherwise''. This modification of the DFA makes it possible to reduce the number of transitions, leaving the number of states of the original DFA intact.

The computational complexity of the Aho-Corasick algorithm directly depends on the data structure. Spatially, the Aho-Corasick FSA with error transitions has $\Sigma_{i=1}^n |w_i|$ states, where $w_i$ is a word from the set of recognizable words, and the number of transitions is reduced from $|A| \times |Q|$ of the original DFA up to $2 \Sigma_{i=1}^n |w_i|$ in the worst case scenario for the   Aho-Corasick FSA with error transitions.

\begin{figure}
	\centering
	\fbox{\includegraphics[width=300pt]{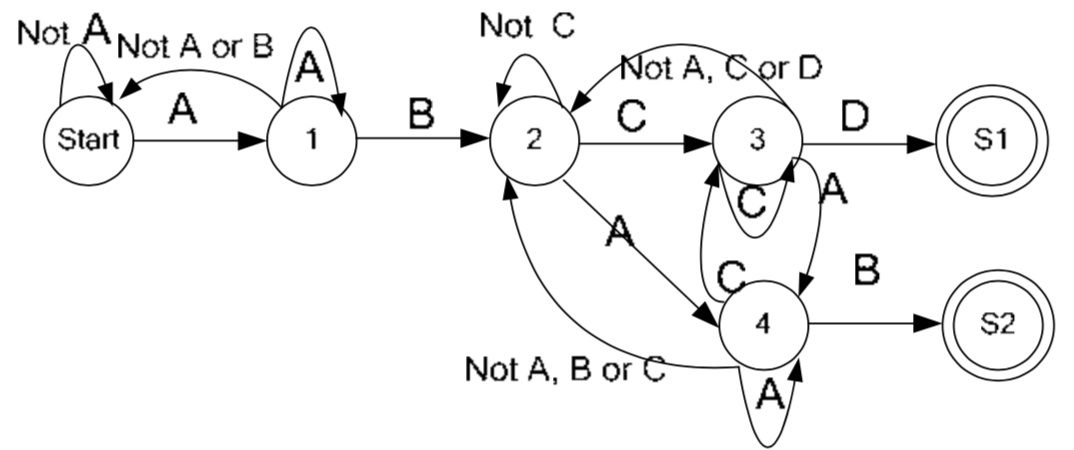}} 

	\caption{The Aho-Corasick based algorithm that uses regular expressions with common prefixes. The modified DFA accepts $.* AB (CD | AB)$ language. Source: ~\cite{Yu.1}}
	\label{fig:fig2}
\end{figure}

An evolution of the Aho-Corasick algorithm is an algorithm that uses regular expressions with common prefixes instead of string~(Fig.~\ref{fig:fig2}) ~\cite{Yu.1}. 

\begin{figure}
	\centering
	\fbox{\includegraphics[width=300pt]{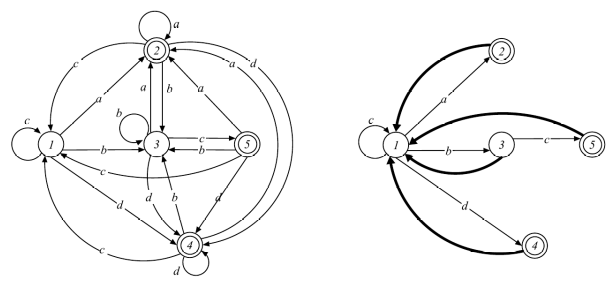}} 

	\caption{DFA and equivalent NFA with delays (D2FA). Source: ~\cite{Kumar.1}}
	\label{fig:fig3}
\end{figure}

However, this evolution to regular expressions was carried out in more detail by Kumar et al. and presented as NFA with delays (D2FA)~\cite{Kumar.1, Kumar.2, Kumar.3}. Like Aho-Corasick NFA with error transitions, NFA with delays allows reducing the transition table without affecting the number of states~(Fig.~\ref{fig:fig3}).

The algorithm for constructing an optimal NFA with delays begins with constructing an undirected graph. Then, a set of spanning trees (using the Kruskal algorithm is constructed from the obtained undirected graphs~\cite{Kruskal.1}. When constructing a D2FA, such a characteristic as the longest path consisting exclusively of default transitions (``the longest $\delta$-path'') is considered.  When choosing an automaton, preference is given to an automaton with the minimum given characteristic from equivalent NFA with delays. The length of the $\delta$-path is one of the parameters that limit the efficiency of compression and the speed of D2FA, so when building a tree, diameter is a restriction. However, building a spanning tree that satisfies the maximum diameter limitation is NP-hard.

The application of this approach to reduce the number of transitions through the construction of a NFA with delays in practice (based on Cisco System 2008, Snort 2008, Bro NIDS 2008) showed that the reduction in the number of transitions could reach 99\% of the initial number of transitions~\cite{Kumar.1, Kumar.2, Kumar.3}.

\begin{figure}
	\centering
	\fbox{\includegraphics[width=300pt]{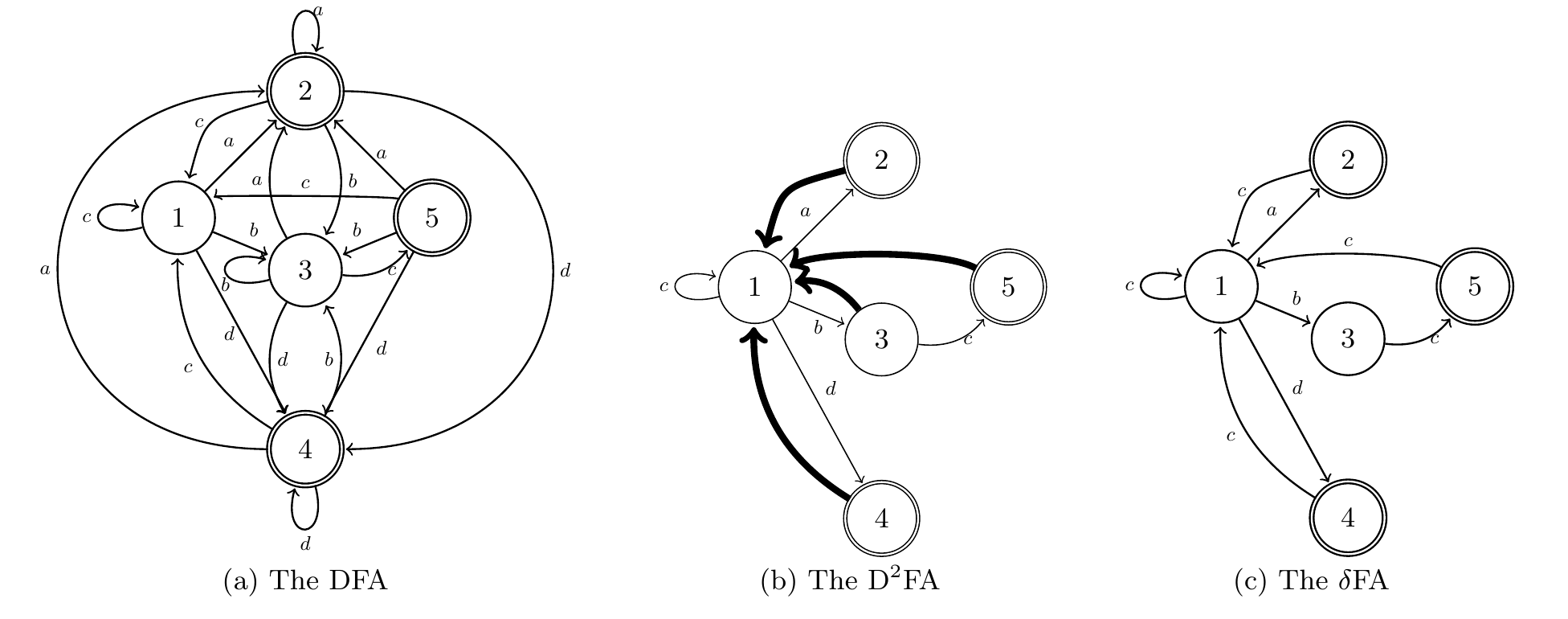}} 
	\fbox{\includegraphics[width=300pt]{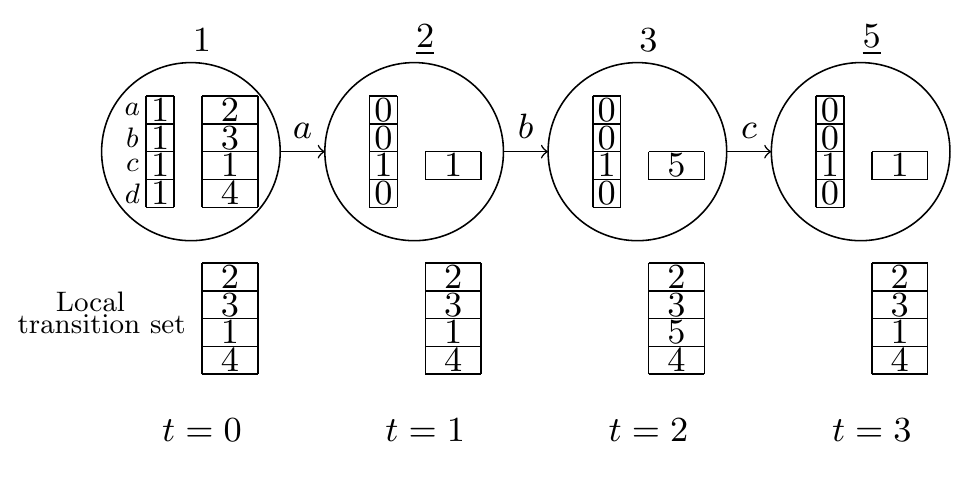}} 
	\caption{$\delta$-FA with the reduced transition table. Source: ~\cite{Ficara.1}}
	\label{fig:fig4}
\end{figure}

A continuation of the idea of NFA with delays (D2FA) is reducing the transition table and, thus, reducing required memory. The idea of Ficara and co-authors of the D2FA modification called $\delta$-FA, is that for each state, not all transition values are stored in the transition table, but only those that differ from the parent state (if any)~(Fig.~\ref{fig:fig4})~\cite{Ficara.1}.

Authors Michela Bechi and Patrick Crowley came up with an algorithm (called ``A Amortized time/bandwidth overhead DFAs'', or A-DFA for short) that exploits the transition redundancy of a recognizing FSA. The redundancy was noticed empirically by the authors and consisted in the fact that most of the states of the DFA, built on data sets of signature databases of intrusion detection systems, had similar sets of outgoing transitions~\cite{Becchi.1}. This was previously noted in the work of Kumar et al.~\cite{Kumar.1, Kumar.2, Kumar.3}. In their work, the A-DFA authors improved the delayed DFA (D2FA) solution proposed by Kumar et al. in 2006~\cite{Kumar.1}. 

The A-DFA algorithm of Michela Becchi and Patrick Crowley uses $2N$ transitions and $N (k + 1)/k$ transitions in the worst case to process a string of length $N$, where $k$ is a predefined value, $k \in \mathbb{N}$~\cite{Becchi.1}. It is shown that, in general, A-DFA provides improved compression over D2FA. As for the time complexity asymptotics, the algorithms for constructing D2FA and A-DFA have time bounds $O(n^{2}\log_2 n)$ and $O(n^{2})$, respectively.

\begin{figure}
	\centering
	\fbox{\includegraphics[width=300pt]{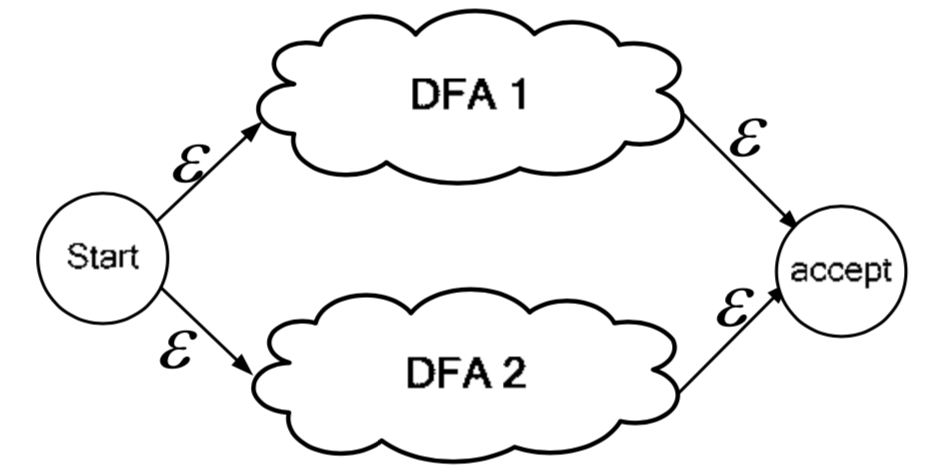}} 
	\caption{Multistride DFA. Source:~\cite{Yu.1}}
	\label{fig:fig11}
\end{figure}

The construction of a multiautomaton (or multistride DFA) is a fundamentally different approach in terms of FSA compression in comparison with the algorithms that were considered above. Compression methods using regular expression grouping are united by the strategy not to adapt to the initial structure of the FSA that recognizes the language, but to select independent (or conditionally independent) subsets in the language. This subsetting allows to build a DFA for each subset and combining the resulting DFAs into a single NFA, while reducing memory requirements~(Fig.~\ref{fig:fig11})~\cite{Yu.1, Rohrer.1, Xu.1}. 

Finding the best memory-efficient grouping of regular expressions is an NP-hard problem. However, the regular expressions grouping for multiautomaton allows multithreading for parallel computing, using each thread for each selected group of interacting regular expressions. 

In practice, in intrusion detection systems and network traffic analysis systems, this algorithm has achieved a memory reduction by order of magnitude compared to the classical DFA~\cite{Yu.1}. Thus, the upper bound on the number of states of such a multiautomatic machine is $O(m2^{l})$, where $m$ is the number of regular expressions $R_i$, from the language $\bigcup^m_{i=1} R_i$, where $l = max(|R_i|)$.

\subsubsection{FSA Structural Modification Approaches}
The most common structural elements that expand the capabilities of the FSA are counters and memory elements that store certain properties of the FSA. This section provides a list of existing solutions, the main idea of which is the transition from an abstract FSA to a structural FSA that combines the abstract part stored in memory and various additives such as bit arrays, circuit, and counters~\cite{Alex.1, Alex.3}. 

In this case, the space complexity is defined as a memory size for the abstract part and the circuit complexity. The exponential number of states is moved into the structural part so that the overall complexity remains acceptable.

The most common structural elements that extend the capabilities of the FSAs are counters and memory elements that store certain grouped states.

\begin{figure}
	\centering
	\fbox{\includegraphics[width=300pt]{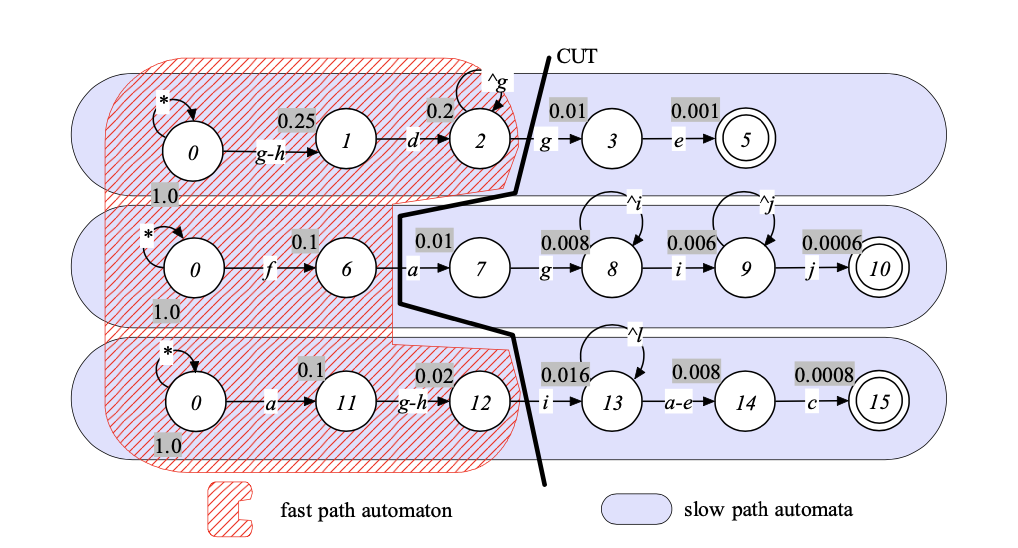}} 
	\caption{DFA with fast and slow memory. Source: ~\cite{Kumar.1} }
	\label{fig:fig5}
\end{figure}

Dealing with malicious traffic recognition, Kumar et al.~\cite{Kumar.1} introduce several concepts that characterize the shortcomings of using the classical DFA in the context of the exponential blow-up problem. Thus, the authors call ``insomnia'' the wasteful storage of all states of the DFA, regardless of the probability of their activation. Indeed, when analysing traffic, most states are rarely activated, and the probability of activating a state decreases with distance from the initial state of the DFA. The authors propose to divide the regular expression into the prefix and suffix parts, determined by the probability of the DFA transition. The authors propose to store transitions to high-probability DFA states in fast memory, while transitions to low-probability states are stored in the ``sleep state'' (slow memory). Switching from fast memory to slow memory should be carried out if necessary by trigger. The trigger is the recognition of the prefix by the fast part of the DFA~(Fig.~\ref{fig:fig5}). 

The peculiarity of such state storage gives a benefit in terms of fast memory consumed while accessing slow memory slows down the operation of this construction compared to the classical DFA. However, a preliminary statistical analysis of traffic can reduce the need to access slow memory. Thus, the construction makes it possible to reduce the memory consumed without significant speed losses~\cite{Kumar.1, Kumar.2}. 

The issue of separating states into probable and improbable ones authors propose to solve empirically by considering the probability distribution of input network traffic. Thus, using the existing database of regular expressions marking malicious traffic is not enough. It is required to "train" the system on the collected dataset, where the traffic will be labeled according to the frequency of states activation. This aspect complicates the use of the solution.

The inability of the FSA to remember ``visited states'' Kumar et al. called ``amnesia'' and suggested storing checking bits in addition to the current state of the FSA~\cite{Kumar.2}. 

\begin{figure}
	\centering
	\fbox{\includegraphics[width=300pt]{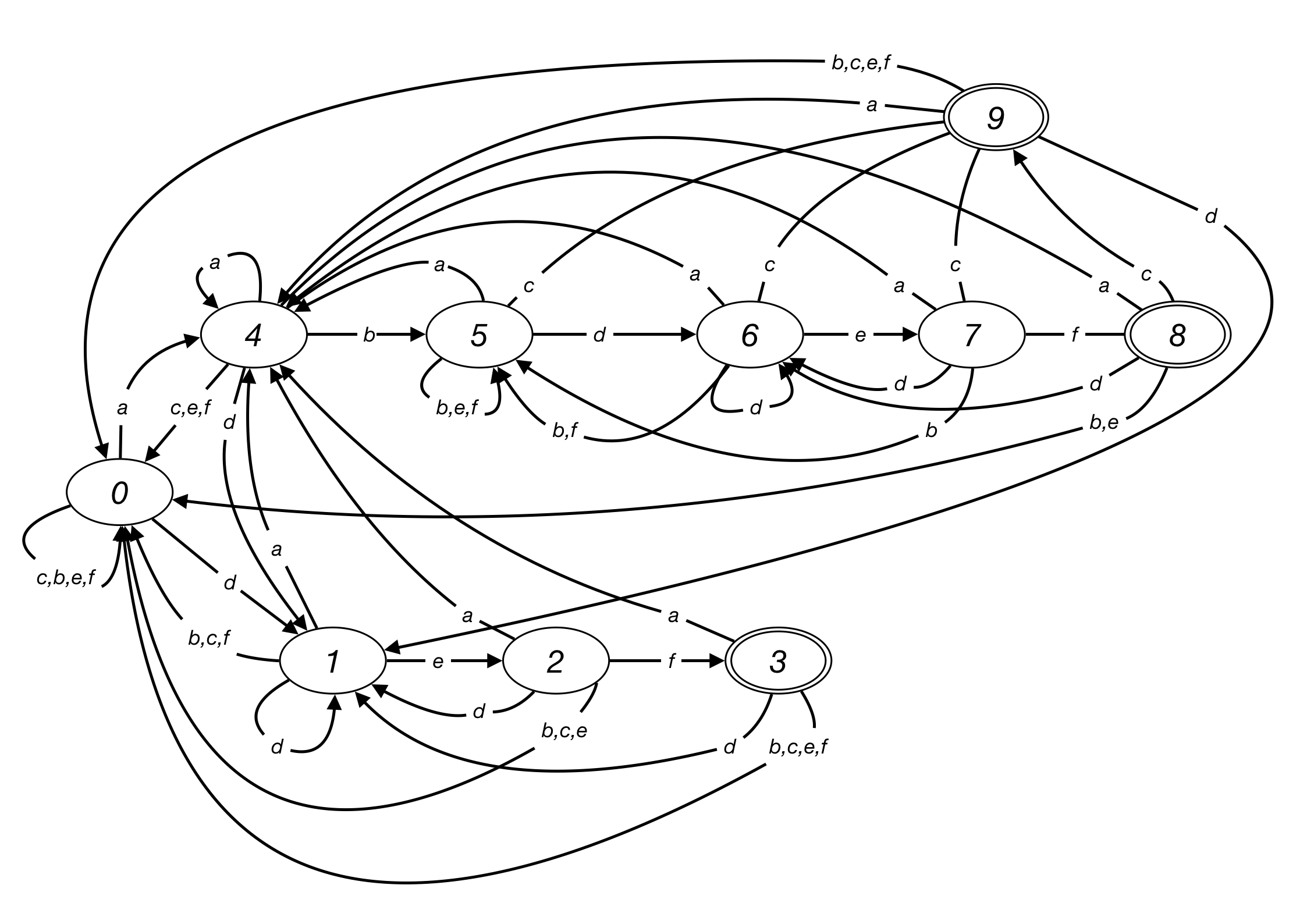}} 
	\fbox{\includegraphics[width=300pt]{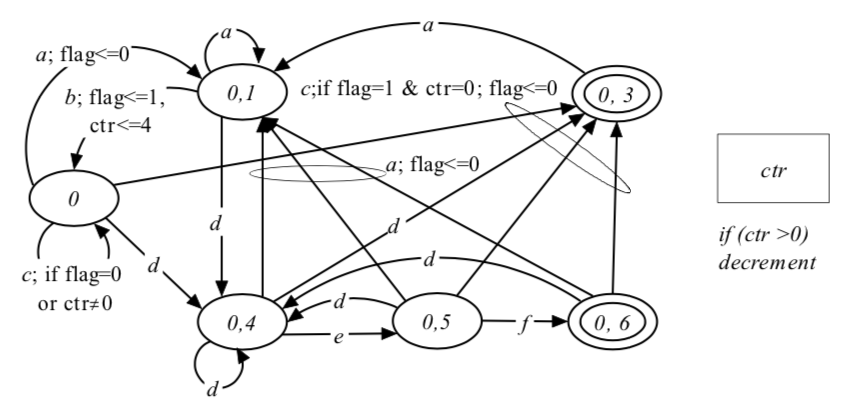}} 
	
	\caption{DFA and equivalent History based Finite Automaton accepting $(.*ab[\neg a]*c) \cup (.*def)$ language. Source: ~\cite{Kumar.2} }
	\label{fig:fig6}
\end{figure}

The automaton with this bits modification is called History based Finite Automaton or H-FA for short~\cite{Kumar.2}. Formally, a DFA with memory is represented by a set, $H-FA = (A, Q, q_0, Q_B, \varphi, H)$, where A is the input alphabet; Q is the alphabet of states; $q_0$ is the initial state; $Q_B$ is the set of accepting states; $H$ is a bit array; $\varphi$ is the transition function, $\varphi : A \times Q \times  H \to Q \times H$. The bit array takes the value equal to 1, if the transition to this state was previously made, and the value equal to 0, if there was no such transition~(Fig.~\ref{fig:fig6}).

\begin{figure}
	\centering
	\fbox{\includegraphics[width=300pt]{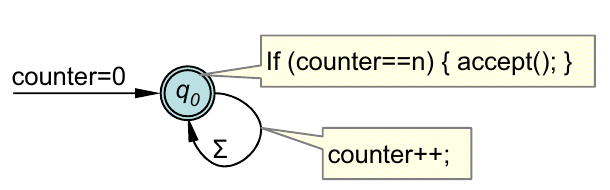}} 
	\caption{The counting structural elements for solving the $\{n\}, \{n, m\}$- related exponential blow-up problem. Source:~\cite{Smith.1}}
	\label{fig:fig7}
\end{figure}

Another Structural Modification Approach is counting structural elements.
Counters complement FSA that matches expressions like $R\{k\}$ and $R\{n, k\}$, where $R$ is a regular expression~(Fig.~\ref{fig:fig7})~\cite{Smith.1}.

The practical application of DFA with memory and counters on existing datasets of intrusion detection systems led to a decrease in the number of states by an order of magnitude compared to the classical DFA~\cite{Kumar.1, Kumar.2, Kumar.3}.

One more Structural Modification Approach is a DFA and NFA combination. The combination of DFA and NFA provides a tangible advantage for solving the exponential blow-up problem by combining the positive properties of determinism and non-determinism in terms of saving resources and time. There are several proposals for using this combination~\cite{Liu.1, Becchi.1}.

Despite many proposed solutions, solutions still need to be improved in front of some regular languages ($(\bigcup_{i=1}^{n}.* R_i) \cup (\bigcup_{i=1}^{n}.* R^{'}_i).*$ language, where $R_i, R^{'}_i$ are regular expressions), as indicated in the section where we described the exponential blow-up problem. The proposed solutions do not make it possible to correctly and quickly check whether a word belongs to a given type of language and run into either inefficiency or the language extension problem described above. In the next section, we provide a universal solution for this type of language.

\section{FSA modification Approach: DFA with counters solves exponential blow-up problem}
\begin{figure}
	\centering
	\fbox{\includegraphics[width=300pt]{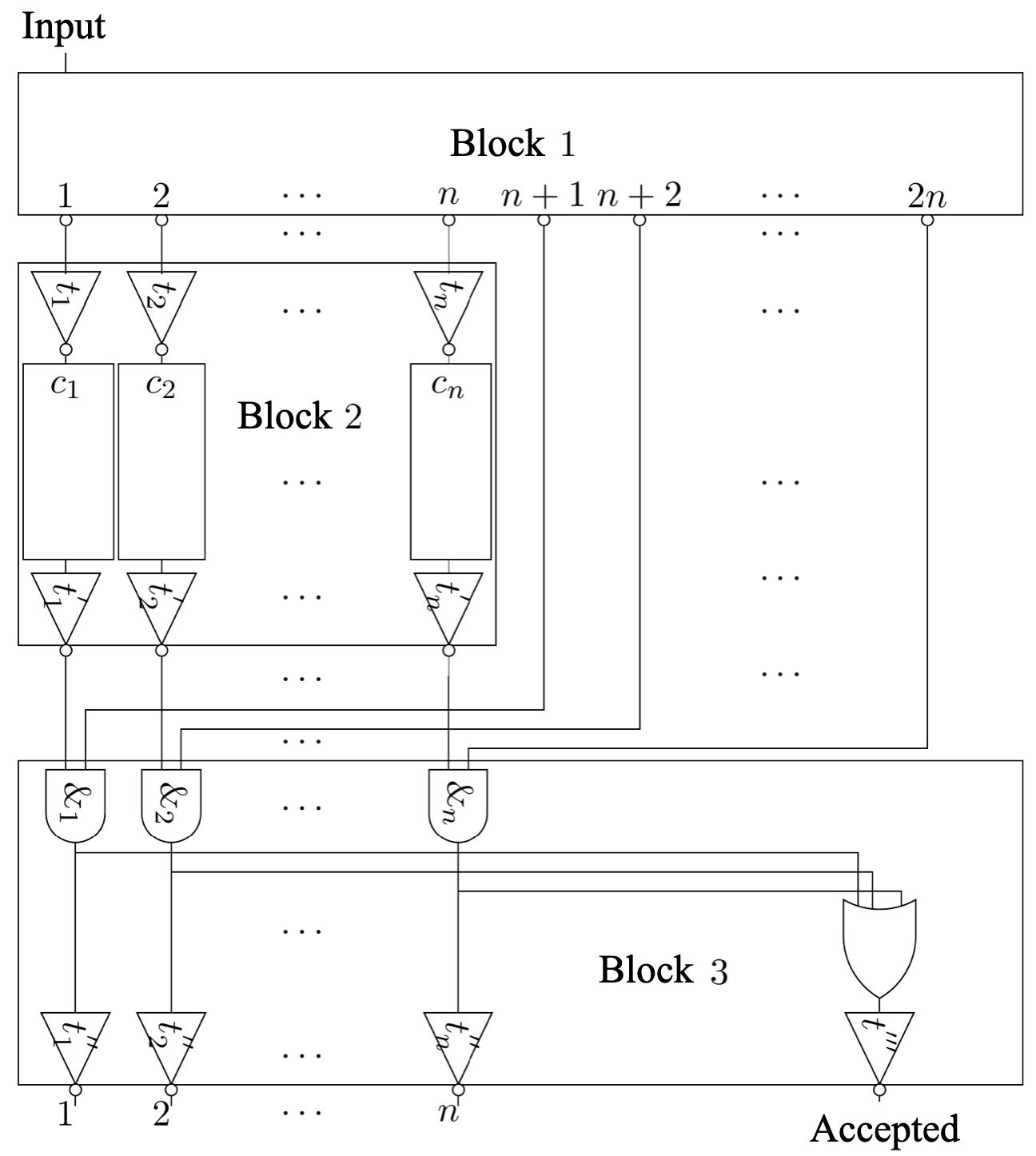}} 
	\caption{DFA with counters construction that accepts the $(\bigcup_{i=1}^{n}.* R_i) \cup (\bigcup_{i=1}^{n}.* \beta_i)$ language, where $R_i$ is a regular expression that does not contain signs that cause the exponential blow-up problem, $\beta_i$, $\gamma_i$ are non-empty words over the alphabet $A$.. 
	Block 1 is a DFA, that accepts the language $(\bigcup_{i=1}^{n}.* R_i) \cup (\bigcup_{i=1}^{n}.* \beta_i)$. In block 2 there are $t_i$, $t_i'$, $t_i''$, $t^{'''}$ triggers and $c_1, ... , c_n$ counters, $i \in \overline{1,n}$ .}
	\label{fig:fig8}
\end{figure}

This part presents the author's solution to the exponential blow-up problem by using a FSA with additions of structural elements~(Fig.~\ref{fig:fig8}).

Presented DFA construction solves the exponential space complexity by decomposing an automaton accepting a given language into two components: an ``abstract'' component, which complexity is defined as the amount of required memory, and a ``structural'' component, which complexity is defined as the number of elements in the schema. The DFA construction accepts the $\bigcup_{i=1}^{n}.* R_i. * \beta_i.*$ regular language, where $R_i$ is a regular expression that does not contain signs that cause the exponential blow-up problem, $\beta_i$ is a non-empty word over the alphabet $A$.

The construction consists of three blocks. Block~1 is a DFA accepting a language $(\bigcup_{i=1}^{n}.* R_i) \cup (\bigcup_{i=1}^{n}.* \beta_i)$, where $R_i$ is a regular expression that does not contain signs that cause the exponential blow-up problem, $\beta_i$ is a non-empty word over the $A$ alphabet. The DFA can be constructed by applying the Aho--Corasick algorithm, memory-efficient regular expressions grouping algorithm, and minimization of automata algorithm~\cite{Aho.1, Brzozowski.1, Kudryavcev.1, Kudryavcev.2}.

If a word from the $(\bigcup_{i=1}^{n}.* R_i)$ language is accepted by the block~1, the counters of the block~2 with preset values are turned on.

The block~2 counts the lengths of the words of the $(\bigcup_{i=1}^{n}.* \beta_i)$ language. It is these counters that ensure that there is no language extension.

The block~3 consists of a layer of $n + 1$ logic gates (n logic gates of the ``conjunction'' type and 1 logic gate of the ``disjunction'' type) and the next layer of $n + 1$ single-pole single-throw (SPST) switches.

If the prefix of a word or a whole word from language $(\bigcup_{i=1}^{n}.* \beta_i)$ is a suffix of a word from language $(\bigcup_{i=1}^{n}.* \alpha_i)$, then it is the counters of the block~2 and logical elements of the block~3 that make it possible not to accept such a word. 

Let us prove the correctness theorem.

\begin{theorem}[the correctness theorem] \label{t1} The presented construction correctly accepts the regular language $\bigcup_{i=1}^{n}.* R_i. * \beta_i.*$, where $R_i$ is a regular expression that does not contain signs that cause the exponential blow-up problem, $\beta_i$ is a non-empty word over the alphabet $A$. The $i$-th output of the construction is equal to 1 if and only if the input word corresponds to the $i-$th summand ($.* R_{i}$), and the $n + 1$-st output is equal to 1 if and only if the word belongs to the $.* R_{i}.*\beta_{i}.*$ language.

\end{theorem}

\begin{proof}[Proof]
Suppose $\gamma$ is a word from the regular language$\bigcup_{i=1}^{n}.* R_i. * \beta_i.*$. Hence $\gamma \in .*R_{i}.*\beta_{i}.*$, $\gamma$ is of the form $\delta \delta^{R}_i \delta^{'} \beta_{i} \delta^{''}$, where $\delta$ does not contain word from $R_i$ as a subword, $\delta^{'}$ does not contain $\beta_i$ as a subword, $\delta^{R}_i$ corresponds to $R_i$. 

This means that after the word $\delta \delta^{R}_i$ is fed to the input, counter number $i$ starts counting, and by the time $\delta \delta^{R}_i \delta^{'}\beta_i$ is finished, firstly, the $i$-th output of the block~2 becomes equal to one, and, secondly, $2n + i$-th output of the block~1 becomes equal to 1. Consequently, the output of the $i$-th conjunction of the block~3 becomes equal to 1 and it will be saved by the $i$-th trigger.

Conversely, let the $i$-th output of block 3 be equal to 1. Consider the moment the value changed from 0 to 1 for the first time. Then at this moment, both inputs of the $i$-th conjunction of block 3 became equal to one. This means that the input word is $\delta^{''}\beta_i$, and at least $|\beta_i |$ steps ago, the word had the form $\delta \delta^{R}_i$. Therefore, the input word is $\delta \delta^{R}_i \delta^{'}\beta_i$ and belongs to the regular language $.* R_{i}.*\beta_{i}.*$. After this moment, the output will remain equal to 1, while any word $.*\delta \delta^{R}_i \delta^{'} \beta_i.*$ also belongs to the language $\bigcup_{i=1}^{n}.* R_{i}.*\beta_{i}.*$.The statement about $n + 1$ output follows from the statement about the first $n$ outputs.

Hence proved.
\end{proof}

\begin{theorem}[theorem on the absence of an exponential blow-up] \label{t2} 
For language recognition, $\bigcup_{i=1}^{n}.* R_i. * \beta_i.*$, where $R_i$ is regular expression that does not contain signs that cause the exponential blow-up problem and $\beta_i$ is non-empty word, the presented construction requires $O(mn (\log_2 m + n))$ bits of memory to store the diagrams of block 1, and $O(n \log_2 m)$ elements to implement blocks 2 and 3, where $m$ is the maximum length of the regular expression $R_i$ and $\beta_i$ word.
\end{theorem}

\begin{proof}[Proof]
Let us look at the first block. The minimal automata algorithm constructs a FSA with the number of states estimated as $О(mn)$, where $m$ is the maximum length of the regular expression $R_i$ and $\beta_i$ word. The transition diagram requires $|A| \times |Q|$ cells of length $] log_2 |Q|[$.

Given that the cardinality of the alphabet $A$ is a constant, block 1 requires $O(mn \log_2 (mn))$ memory bits to store the transition diagram. 

The output has $2n$ dimension; therefore, it is needed $|A| \times |Q|$ cells of length $2n$ to store the output function diagram. That is, $O(mn^2)$. Altogether, for both diagrams it requires $O(mn (\log_2 (mn) + n))$ memory bits or $O(mn \log_2 m + mn^{2})$. Thus, block 1 requires $O(mn (\log_2 m + n))$ memory bits to store the diagrams. 

Let us look at the second block. It consists of a linear in $n$ number of triggers and counters. The trigger has constant complexity; the counter is logarithmic in $m$, where $m$ is the maximum length of the regular expression $R_i$ and $\beta_i$ word. As a result, we get the complexity $O(n \log_2 m)$.

Finally, the third block is linear in $n$ and constant in $m$. Thus, 2 and 3 blocks have the structural complexity of order $O(n \log_2 m)$.

Hence proved.

\end{proof}
\subsection{Multi extension of the DFA with counters}

\begin{figure}
	\centering
	\fbox{\includegraphics[width=300pt]{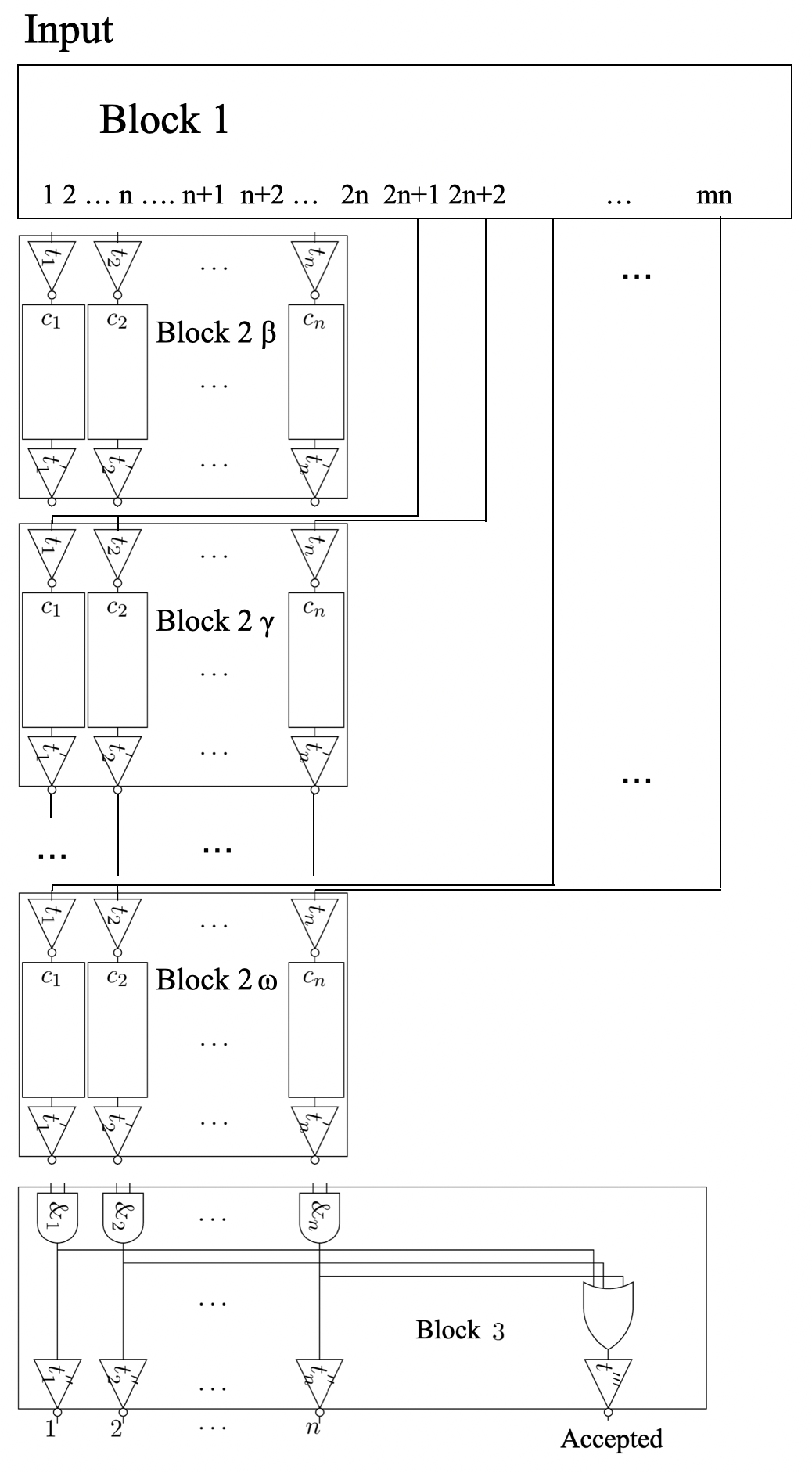}} 
	\caption{DFA with counters construction, that accepts the $\bigcup_{i=1}^{n}.* R_i. * \beta_i.* \gamma_i .* ... \omega_i .*$ language, where $R_i$ is a regular expression that does not contain signs that cause the exponential blow-up problem, $\beta_i$, $\gamma_i$, ... ,$\omega_i$ are non-empty words over the alphabet $A$. 
	Block 1 is a DFA, that accepts the language $(\bigcup_{i=1}^{n} .* R_i ) \cup (\bigcup_{i=1}^{n} .* \beta_i ) \cup (\bigcup_{i=1}^{n} .* \gamma_i ) \cup ... \cup (\bigcup_{i=1}^{n} .* \omega_i)$. In block 2 there are $t_i$, $t_i'$, $t_i''$, $t^{'''}$ triggers and $c_1, ... , c_n$ counters, $i \in \overline{1,n}$ .}
	\label{fig:fig9}
\end{figure}

The construct naturally extends to the following languages: $\bigcup_{i=1}^{n}.* R_i. * \beta_i.* \gamma_i .* ... \omega_i .*$, where $R_i$ is a regular expression that does not contain signs that cause the exponential blow-up problem, $\beta_i, \gamma_i, ..., \omega_i$ are non-empty words over the alphabet $A$. The modified block 1 accepts $(\bigcup_{i=1}^{n} .* R_i ) \cup (\bigcup_{i=1}^{n} .* \beta_i ) \cup (\bigcup_{i=1}^{n} .* \gamma_i ) \cup ... \cup (\bigcup_{i=1}^{n} .* \omega_i)$, where $R_i$ is a regular expression that does not contain signs that cause the exponential blow-up problem, $\beta_i$, $\gamma_i$, ... ,$\omega_i$ are non-empty words over the alphabet $A$. For each, except for the first $\bigcup_{i=1}^{n}.* R_i$, its own block 2 with counters is created~(Fig.~\ref{fig:fig9}).

\subsection{Quantifier extension of the DFA with counters}
\begin{figure}
	\centering
	\fbox{\includegraphics[width=300pt]{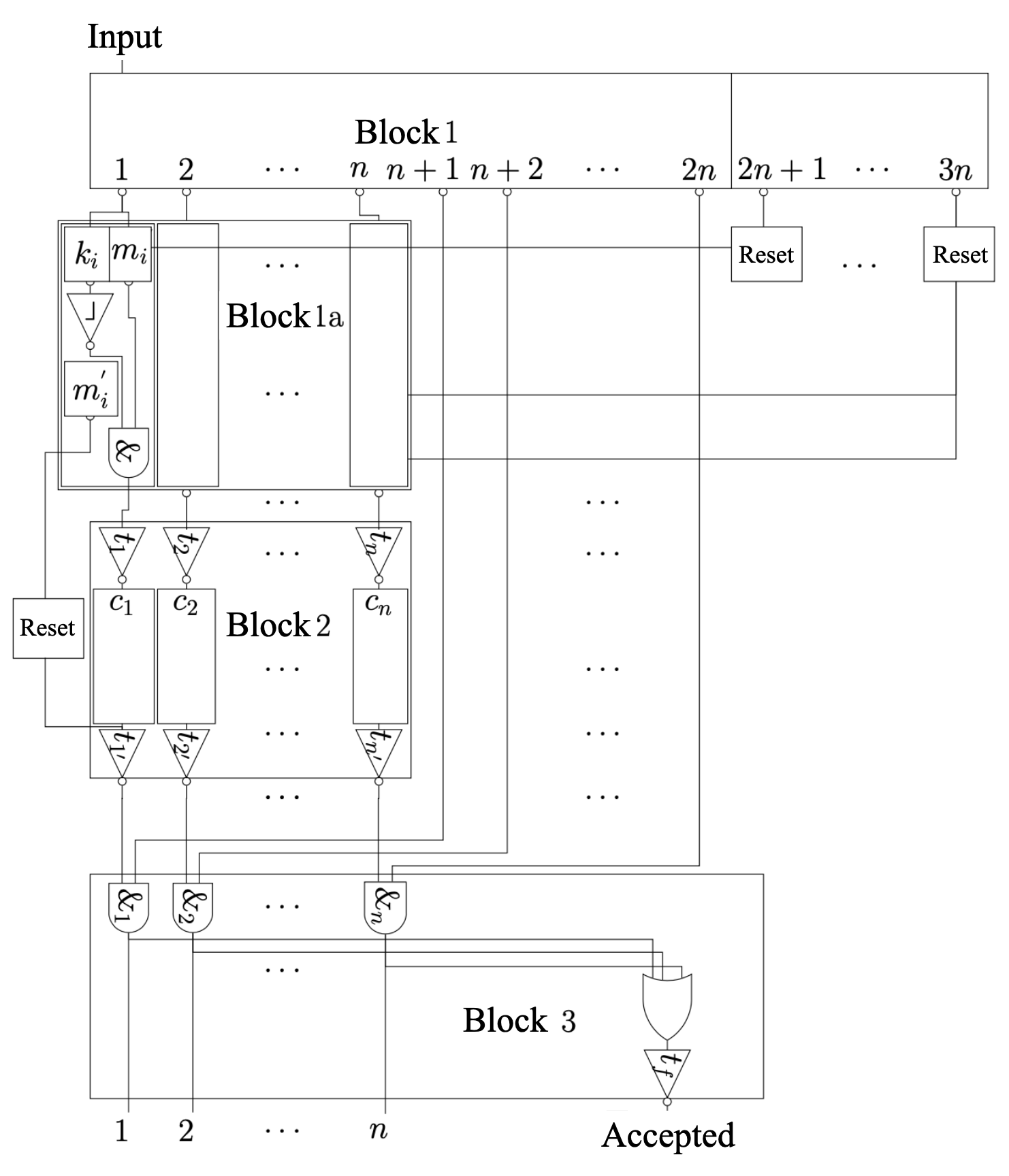}} 
	\caption{The extended DFA with counters construction. Block 1 is a DFA that accepts the language $(\bigcup_{i=1}^{n} .* R_i ) \cup (\bigcup_{i=1}^{n} .* \beta_i ) \cup (\bigcup_{i=1}^{n} .* \gamma_i )$, where $R_i$ is a regular expression that does not contain signs that cause the exponential blow-up problem, $\beta_i$, $\gamma_i$ are non-empty words over the alphabet $A$. Block 1a is a counting block. $t_i$, $t_i'$, $t_i''$, $t^{'''}$ are triggers, $i \in \overline{1,n}$  $c_1, ... , c_n$ are counters in a counting block 2.}
	\label{fig:fig10}
\end{figure}

The extended construction accepts $\bigcup_{i=1}^{n}.* R_i [ \neg \gamma_i] \{k_i, m_i\} \beta_i.*$ and $\bigcup_{i=1}^{n}.* R_i [ \neg \gamma_i] \{k_i\} \beta_i.*$ languages, where $R_i$ is a regular expression that does not contain signs that cause the exponential blow-up problem and $\beta_i$, $\gamma_i$ are non-empty words over the alphabet $A$, the values in the curly brackets set the number of repetitions (for $\{k_i\}$) and the interval $\{k_i, m_i\}$ of repetitions of the words other than $\gamma_i$~(See Fig.~\ref{fig:fig10}). 

In the extended construction, block 1 accepts the language $(\bigcup_{i=1}^{n} .* R_i ) \cup (\bigcup_{i=1}^{n} .* \beta_i ) \cup (\bigcup_{i=1}^{n} .* \gamma_i )$, where $R_i$ is a regular expression that does not contain signs that cause the exponential blow-up problem, $\beta_i$, $\gamma_i$ are non-empty words over the alphabet $A$.

From the $\overline{1, n}$ outputs of the block~1, the signal goes to the block~1a corresponding to the $\alpha_i$. In the block~1a, counters with preset values $k_i$, $m_i$, $m_i^{'}$ are switched on, where $k_i$, $m_i$ corresponds to the values in curly brackets of the $.* R_i[\neg \gamma_i]{k_i, m_i}.* \beta_i .*$, and $m_i^{'}$ corresponds to the value $(m_i + |\beta_i| + 1)$. The $i$-th input of the block~1a is switched on to receive a signal from the $2n + i$-th output of the block~1.

The block~2 counts the lengths of the words of the $\bigcup_{i=1}^{n} .* \beta_i$. 
The block~3 consists of a layer of $n + 1$ logic gates ($n$ ``conjunctions'' and 1 ``disjunction'') and the next layer of the $n + 1$ SPST switches.


\subsection{Double counting extension of the DFA with counters}
The proposed DFA with counters design can be additionally equipped with counting structural elements in block 1~(Fig.~\ref{fig:fig8}). For example, counters that were proposed in the work of colleagues to avoid the exponential blow-up of the $\{n\}, \{n, m\}$ regular signatures~\cite{Smith.1}.

This double counting extension allows solving the exponential explosion problem for a class of languages, such as:
\begin{equation}
\Bigg( \bigcup_{i=1}^{n}.* R_i. * \beta_i.* \gamma_i .* ... \omega_i .* \Bigg) \bigcup \Bigg( \bigcup_{i=1}^{n}.* R_i [ \neg \gamma_i] \{k_i, m_i\} \beta_i.*  .* \gamma_i .* ... \omega_i .* \Bigg),
\end{equation}
where $R_i$ is a regular expression that does not contain wild-card signs ``.*'' and ``.+'' and can contain $\{n\}, \{n, m\}$ counters, $\beta_i,  \gamma_i, ..., \omega_i$ are non-empty word over the alphabet $A$, $n, k, m \in \mathbb {N}$, $i \in \overline{1, n}$.

Moreover, the theorem~\ref{t2} on the absence of an exponential blow-up remains true.

Let us look at the first block. The block 1 requires $O(mn (\log_2 m + n))$ memory bits to store the DFA diagrams and a linear in $n$ number of triggers and counters with $O(n \log_2 m)$ structural complexity, where $m$ is the maximum length of the regular expression $R_i$ and $max ( |\beta_i|, |\gamma_i|, ... |\omega_i|)$ non-empty words.

 Altogether, block 1 requires $O(mn (\log_2 m + n))$ elements.

\section{Conclusion}

In solving the regular expression matching problem, some ideas and algorithms are exceptionally successful, which became the progenitors of entire directions in signature analysis. 

First, such findings include the Aho-Corasick searching algorithm with error transitions. The algorithms can significantly reduce the memory requirements in practice. However, the algorithms that use transitions in case of an error do not solve the exponential blow-up problem in the general case.

Secondly, building a multi-automatic machine based on a grouping of interacting regular expressions initiated parallel computing for regular expression matching. However, despite a decent number of ideas, regular expression grouping algorithms remain an NP-complete problem and do not guarantee a solution to the exponential blow-up problem in the general case. 

Despite many proposed solutions, solutions still need to be improved in front of some regular languages (such as $\bigg( \bigcup_{i=1}^{n}.* R_i. * \beta_i.* \gamma_i .* ... \omega_i .*\bigg) \bigcup \bigg( \bigcup_{i=1}^{n}.* R_i [ \neg \gamma_i] \{k_i, m_i\} \beta_i.*  .* \gamma_i .* ... \omega_i .* \bigg)$, where $R_i$ are regular expressions, $\beta_i,  \gamma_i, ..., \omega_i$ are non-empty word over the alphabet $A$, $n, k, m \in \mathbb {N}$, $i \in \overline{1, n}$), which cause the exponential blow-up problem. 

This paper provides a theoretical and hardware solution to the exponential blow-up problem for the complicated classes of regular languages, which caused severe limitations in Network Intrusion Detection Systems work. 

The article supports the solution with theorems on correctness and complexity. 

\bibliographystyle{unsrtnat}


\end{document}